\documentclass[11pt,a4paper]{article}
\usepackage{amsmath}
\usepackage{amsfonts}
\usepackage{amssymb}
\usepackage{graphicx}
\usepackage[left=1.00in, right=1.00in]{geometry}
\usepackage{multirow}

\usepackage{setspace}
\usepackage{hyperref}
\usepackage[round]{natbib}

\usepackage{enumerate}

\usepackage{rotating}

\newcommand{\expect}{\mathbb{E}}
\DeclareMathOperator{\rank}{rank}

\graphicspath{{./figures/}}

\begin{document}

\title{Evaluating tests for cluster-randomized trials with few clusters under generalized linear mixed models with covariate adjustment: a simulation study}
\author{Hongxiang Qiu, Andrea J. Cook, Jennifer F. Bobb}
\date{}

\maketitle

\begin{abstract}
	Generalized linear mixed models (GLMM) are commonly used to analyze clustered data, but when the number of clusters is small to moderate, standard statistical tests may produce elevated type I error rates. Small-sample corrections have been proposed for continuous or binary outcomes without covariate adjustment. However, appropriate tests to use for count outcomes or under covariate-adjusted models remains unknown. An important setting in which this issue arises is in cluster-randomized trials (CRTs). Because many CRTs have just a few clusters (e.g., clinics or health systems), covariate adjustment is particularly critical to address potential chance imbalance and/or low power (e.g., adjustment following stratified randomization or for the baseline value of the outcome). We conducted simulations to evaluate GLMM-based tests of the treatment effect that account for the small {(10) or moderate (20) number} of clusters under a parallel-group CRT setting across scenarios of covariate adjustment (including adjustment for one or more person-level or cluster-level covariates) for both binary and count outcomes. We find that when the intraclass correlation is non-negligible {($\geq 0.01$)} and the number of covariates is small {($\leq 2$)}, likelihood ratio tests with a between-within denominator degree of freedom have type I error rates close to the nominal level. When the number of covariates is moderate {($\geq 5$)}, across our simulation scenarios, the relative performance of the tests varied considerably and no method performed uniformly well. 
Therefore, we recommend adjusting for no more than a few covariates and using likelihood ratio tests with a between-within denominator degree of freedom.
\end{abstract}

\section{Introduction}

Generalized linear mixed models (GLMM) are commonly used to analyze clustered data. However, when the number of clusters is small to moderate, standard tests of model coefficients may be seriously anticonservative \citep{kahan2016}. To address inflated type I error rates due to having a small number of clusters, prior literature has proposed and validated small-sample corrected testing procedures for linear mixed models (LMMs) and logistic generalized linear mixed models (GLMMs) \citep{schluchter1990,schaalje2002,pinheiro2006,li2015,mcneish2016,leyrat2018,bolker2009,vonesh2014,thompson2022}. The general approach of these studies is to conduct an ordinary t-test or F-test but with a carefully chosen denominator degree of freedom (DDF) that takes the small number of clusters into account. For example, the Kenward-Roger DDF has been found to perform relatively well in linear mixed models when the data consist of repeated measures with a complicated covariance structure, though it still has elevated type I error rates \citep{schaalje2002}; Wald F tests with the ``between-within'' DDF have been shown to perform well for continuous \citep{leyrat2018} and binary outcomes \citep{li2015}. These studies focused on Wald tests, whereas likelihood ratio tests (LRTs) were reported to be unreliable with a small number of clusters \citep{bolker2009,vonesh2014,li2015}. Despite this rich literature, key gaps in knowledge remain on which method achieves type I error rates close to the nominal level. One gap is the choice of tests when analyzing count outcome data with Poisson GLMMs. In addition, it is common to adjust for covariates in analyses, but small-sample procedures have not, to our knowledge, been evaluated previously under covariate-adjusted models.

An important area of application where such issues arise is cluster-randomized trials (CRTs). Many CRTs have only a few clusters because the intervention is at the cluster-level and may be expensive. For example, two reviews of cluster-randomized trials (CRTs) found the median number of clusters to be 21 \citep{eldridge2004} and 25 \citep{kahan2016}. Having few clusters may lead to imbalance in key prognostic variables between arms due to chance, which can result in misleading results that are sometimes termed chance bias \citep{roberts1999}. In addition, unadjusted analyses may have low power. To alleviate these issues, it is common to adjust for one or more covariates in the statistical analysis. For example, some trials conduct a stratified randomization and then adjust for the cluster-level stratification variables in the analysis. It is also often desirable to adjust for the baseline measure of the outcome and/or other prognostic cluster-level and person-level covariates to improve power \citep{kahan2014}. One review found that covariates were used in randomization of 58\% of trials, and among trials where baseline measures of the primary outcome were reported, 75\% subsequently adjusted for these covariates in analyses \citep{wright2015}. Consequently, because GLMM analyses in CRTs are often challenged by having a small number of clusters and involve covariate adjustment, knowledge of the correct test to use in this setting is critical for obtaining valid inference in many studies.

In this paper, we investigate the performance of small-sample correction methods for GLMM via simulation, with a focus on testing treatment effects on binary and count outcomes in covariate-adjusted analyses. We consider a parallel group CRT setting, where each cluster is randomized to a treatment arm and then person-level outcomes are measured. Our simulation scenarios cover common settings of covariate adjustment: (1) adjusting for a single cluster-level covariate (corresponding to covariate adjustment following stratified randomization), (2) adjusting for a single person-level covariate (corresponding to adjusting for the baseline value of the outcome), and (3) adjusting for a larger number of person-level and cluster-level covariates (corresponding to studies with several important predictors of the outcome). {We also illustrate these methods with an application to the PRimary Care Opioid Use Disorders Treatment (PROUD) trial, a parallel two-group cluster-randomized implementation trial randomizing 12 clinics across 6 health care systems \citep{PROUDprotocol,PROUDresult}}.

\section{Methods}

\subsection{Problem setup}

We consider a scenario of conducting a parallel group CRT and analyzing the trial data with a generalized linear mixed model (GLMM). Suppose there are $K$ clusters, each with $n_i$ ($i=1,\ldots,K$) persons. To statistically infer the treatment effect, we consider fitting a GLMM that has link function $g$ and takes the form
$$g(\expect[Y_{ij} | \gamma_i]) = X_{ij}^\top \beta + Z_i \gamma_i \quad (i=1,\ldots,K,j=1,\ldots,n_i).$$
where $\gamma_i \sim N(0,\tau^2)$ and $\beta$ is the vector of fixed effects. Here, for each cluster $i=1,\ldots,K$ and person $j=1,\ldots,n_i$, we use $Y_{ij}$ to denote the outcome of interest, $X_{ij}$ to denote the vector containing the variables for which fixed effects are included in the model, and $\gamma_i$ to denote the random intercept of cluster $i$. We note that $X_{ij}$ includes a covariate being the treatment indicator, whose associated fixed effect is of interest; it may also include other covariates, either at person- or cluster-level, to be adjusted for in the model. Although in principle, adjusting for other covariates is unnecessary to obtain a consistent estimator of (and an asymptotically valid confidence interval for) the treatment effect, efficiency and power may be improved if covariates such as the baseline outcome are adjusted for. Moreover, covariate adjustment can be important in CRTs with a small number of clusters, because it can mitigate chance imbalance between treatment arms. The collection of $Z_i$ forms the design matrix $Z$ for random effects; the collection of $X_{ij}$ forms the design matrix $X$ for fixed effects. We further let $N := \sum_{i=1}^{K} n_i$ be the total number of persons.

\subsection{Methods compared}

There are several proposed denominator degrees of freedom (DDF) for Wald t-test or likelihood ratio F-test (LRT) when conducting inference about a slope parameter in GLMMs. In this work, we consider the DDFs that can be easily applied across statistical software packages commonly used in the biomedical research (R, SAS, and Stata). In the subsequent simulations and data analyses, we used the R package \textsf{lme4} \citep{lme4package} to fit GLMMs. The method to integrate random effects is the default method, Laplace approximation.

\subsubsection{Residual}

The residual DDF does not account for the small number of clusters. It takes the form of the DDF in the corresponding linear model without accounting for clustering; that is, the residual DDF equals $N-\rank(X)$, where $X$ is the design matrix of fixed effects.

\subsubsection{Containment}

The containment DDF \citep{li2015} takes the number of clusters into account. It is computed as follows: first search for all random effects that contain the covariate whose fixed effect is of interest, then compute the rank contribution to $\rank([X \,\, Z])$ of each random effect, and finally choose the smallest among the rank contributions; when no such random effect exists, the containment DDF is computed as $N-\rank([X \,\, Z])$. Under our framework of CRTs, when the treatment effect is the parameter of interest, since this covariate is not contained in any random effect, the containment DDF is computed as $N-\rank([X \,\, Z])$. Therefore, although containment DDF takes the number of clusters into account, when $N$ is much greater than $K$, we can expect that the containment DDF is still large and performs similarly to the residual DDF.

In many CRTs with a small number of clusters, the total number of individuals is large. In such cases, both containment and residual DDFs are large, leading to an approximate normal or $\chi^2$ reference distribution in the test. Consequently, these two DDFs might account for the small number of clusters only to a limited extent.

\subsubsection{Between-within}

The between-within (BW) DDF was introduced under the context of longitudinal data and repeated measures \citep{schluchter1990}. It has also been investigated under the context of analyzing a CRT without adjusting for any covariates via simulation \citep{li2015}. In this previous study, the formula of BW DDF is $K-\rank(X)$. However, the authors did not focus on the scenario of covariate adjustment. There can be two natural generalizations of this formula to the setting with covariate adjustment, which we will call BW1 and BW2 respectively for the rest of this article.

The approach in BW1 DDF is to directly use the same formula $K-\rank(X)$ with $X$ being the design matrix of fixed effects. This is also how the authors defined BW DDF for the general GLMM in this previous study \citep{li2015}. Although this generalized BW DDF is straightforward to calculate, in an analysis where a large number of person-level covariates are adjusted, it is possible that $\rank(X) \geq K$ and consequently BW1 DDF is undefined. However, it is not uncommon to adjust for person-level characteristics such as age, race/ethnicity and sex, which already contribute at least three to $\rank(X)$.

BW2 DDF, also called the inner-outer DDF \citep{pinheiro2006}, is another generalization. Its definition may seem complicated, but the intuition is to split the residual degrees of freedom into between-cluster and within-cluster portions, and for each fixed effect, use the within-cluster degrees of freedom if the associated covariate varies within a cluster, and otherwise, use the between-cluster degrees of freedom. In our framework of CRTs, the BW2 DDF is $K$ minus the rank of the design matrix of the fixed effects associated with the cluster-level covariates and further minus one if the intercept is included in the model. Although BW2 DDF is still undefined if a large number of cluster-level covariates is adjusted for, in CRTs, this scenario may be rare and it is more common to adjust for a large number of person-level covariates. We also note that BW1 DDF is always less than or equal to BW2 DDF, and hence a test using BW1 DDF is no less conservative than the corresponding test using BW2 DDF.

\subsection{Simulation}

We conduct a simulation study to investigate the type I error rate of the aforementioned testing methods in a CRT setting with a small number of clusters. We vary the number of clusters, the mean cluster size, the coefficient of variation (CV) of cluster sizes and the intraclass coefficient (ICC). We consider data generating models with various numbers of person- and cluster-level covariates. We also consider fitted models that may include redundant person- or cluster-level covariates in order to study the possible scenario where covariates that are not strongly predictive of the outcome are adjusted for.

\subsubsection{Data generating mechanism in simulation}

We conduct simulation for binary and count outcomes with logistic and Poisson GLMM separately. We consider the following scenarios for cluster sizes: the number of clusters may be small (10) or moderate (20); the average cluster size may be moderate (50) or large (100); the CV (coefficient of variation) of cluster sizes may be 0, 0.75 or 1.5. The CV is the ratio of the standard deviation of cluster sizes to the mean cluster size and is a measure of the variability of cluster sizes. A larger CV indicates more substantial variability, and when CV=0, all clusters have equal sizes.

To generate the cluster sizes, we consider the following approach that can match the mean cluster size and CV with the value of the given scenario while ensuring that each cluster has at least five persons. Specifically, for a given mean cluster size $s$ and CV, we generate a negative binomial random variable $R$ with mean being $s-5$ and variance being $(s \cdot \text{CV})^2$, and then take the cluster size to be $R+5$.

The covariates are generated as follows. We generate two univariate person-level covariates, $X^{\text{person,1}}_{ij}$ and $X^{\text{person,2}}_{ij}$, independently from a standard normal, and two univariate cluster-level covariates, $X^{\text{cluster,1}}_i$ and $X^{\text{cluster,2}}_i$, independently from Bernoulli(0.5). We distinguish covariates at these two levels because the level of each covariate affects the computation of containment and BW2 DDFs. The first covariate at each level may be predictive of the outcome while the second covariate is redundant (i.e., not associated with the outcome) and is adjusted for in some fitted models. We consider this redundant covariate because in practice, some covariates might not be highly predictive but still included in a prespecified model in the statistical analysis plan. We then generate the treatment assignment stratified by below vs.~above the median of the generated cluster sizes, with each arm having equal number of clusters. The purpose of stratification is to balance the cluster sizes in both arms.

To generate the outcome, we consider the following general model:
$$g(\expect[Y_{ij}|\gamma_{i}])=\beta^0 + \beta^{\text{person}} X^{\text{person,1}}_{ij} + \beta^{\text{cluster}} X^{\text{cluster,1}}_i + \gamma_{i}$$
with $\beta^0=0$ and the link function $g$ is the logistic link when the outcome is binary and the log link when the outcome is a count. We consider four settings of this general model in terms of which variables are predictive of the outcome:
\begin{enumerate}[(A)]
	\item None: $\beta^{\text{person}}=\beta^{\text{cluster}}=0$;
	\item The person-level covariate: $\beta^{\text{person}}=0.7$, $\beta^{\text{cluster}}=0$;
	\item The cluster-level covariate: $\beta^{\text{person}}=0$, $\beta^{\text{cluster}}=0.8$;
	\item Both person- and  cluster-level covariates: $\beta^{\text{person}}=0.7$, $\beta^{\text{cluster}}=0.8$.
\end{enumerate}
Note that these models do not include a treatment effect term because we aim to investigate type I error rates. After generating conditional means for each person using the above mean model, we generate the outcomes from a Bernoulli or Poisson distribution with the corresponding mean, depending on whether the outcome is a binary or count measure.

We consider five levels of ICC: 0.001, 0.01, 0.05, 0.1 and 0.2. We use the following definition of ICC \citep{nakagawa2017ICC}: for binary outcomes, the ICC is defined as $\tau^2/(\tau^2+\pi^2/3)$, which uses the theoretical person-level variance specific to logistic models; for count outcomes, the ICC is defined as $\tau^2/\{\tau^2+\log[1+1/\expect(Y_{ij})]\}$, which uses lognormal approximation to person-level variance. Note that in this definition, the random effect and covariates are marginalized in the expectation. The variance of the random effect $\gamma_{i}$, $\tau^2$, is then determined by the level of ICC. Because there are various definitions of ICC for clustered data, among which those based on LMMs are popular, we also report the mean ICC based on LMMs (LMM ICC) \citep{donner1980,stanish1983} that adjust for covariates in the data-generating model. For each generated data set, the LMM ICC is the ratio of the estimated random effect variance to the estimated total variance from the LMM with covariates adjusted for. We take the average of these LMM ICCs across simulated data sets as the mean LMM ICC. As shown in Sections~\ref{section: results sim1} and \ref{section: results sim2}, the performance of the methods highly depends on the ICC. Therefore, we report the mean LMM ICC as an additional anchor that may guide future studies in selecting an analysis method given the within-cluster correlation.

\subsubsection{Investigated statistical procedures}

In each simulation setting, we vary the set of covariates being adjusted for in the fitted model. Specifically, we consider models with the following sets of adjusted redundant covariates in addition to the covariates in the data-generating model:
\begin{enumerate}[(1)]
	\item None;
	\item $X^{\text{person,2}}_{ij}$;
	\item $X^{\text{cluster,2}}_{i}$;
	\item $X^{\text{person,2}}_{ij}$ and $X^{\text{cluster,2}}_{i}$.
\end{enumerate}
As discussed above, we consider these fitted models because, in practice, study teams may not know whether all included covariates will truly be prognostic in their particular setting. For each model, we consider testing the treatment effect with Wald t-test or LRT with four choices of DDF: residual, containment, BW1 and BW2. The nominal type I error rate is fixed at 0.05. For each simulation setting, we simulate 5000 runs and obtain the type I error rates of these tests estimated via Monte Carlo.

At most two person-level covariates are adjusted for in this simulation. Particular scenarios of practical relevance include these combinations of data-generating and fitted models: (i) (A, 1) corresponding to no covariate adjustment, (ii) (B, 1) corresponding to adjusting for the baseline outcome, which is often a prognostic covariate, (iii) (C, 1) or (A, 3) corresponding to adjusting for a potentially prognostic cluster-level covariate, and (iv) (D, 1) or (B, 3) corresponding to adjusting for both the baseline outcome and a potentially prognostic cluster-level covariate. An example of a potentially prognostic cluster-level covariate is one on which randomization is stratified, but we note that in our simulation randomization is not stratified on such prognostic covariates. Although our data generation stratified randomization on cluster size (to improve efficiency by balancing sample size across arms), because cluster size was not considered as a prognostic variable, we do not adjust for it in the fitted models since in our simulation scenarios cluster size was not related to the outcome. In practice, adjusting for stratification variables is important if such factors are associated with the outcome, though a philosophical argument can be made that all factors used to stratify randomization should be adjusted for (regardless of whether they are known to be prognostic). The above combinations are of particular interest because primary analyses often correspond to one of these cases. We also note that BW1 and BW2 coincide when the covariates being adjusted for are all at the cluster-level, i.e., for these combinations of data-generating and fitted models: (A, 1), (A, 3), (C, 1) and (C, 3). 

\subsubsection{Second simulation with greater difference between BW1 and BW2 DDFs}

In the previous simulation setting, BW1 and BW2 DDFs differ by at most two because there are at most two person-level covariates to be adjusted for. In some applications, the number of person-level covariates being adjusted for might be much larger, which may even make BW1 DDF undefined. To better reflect this scenario, we conduct another simulation that is similar to the previous one but with more person-level covariates. We do not vary the number of cluster-level covariates because BW1 and BW2 only differ in how they handle person-level covariates.

In this simulation, we generate four univariate person-level covariates independently as follows:
$$X^{\text{person,1}}_{ij} \sim N(0,1), \, X^{\text{person,2}}_{ij} \sim N(0,1), \,
X^{\text{person,3}}_{ij} \sim \text{Bernoulli}(0.5), \, X^{\text{person,4}}_{ij} \sim N(0,1).$$
As in the first simulation, the last person-level covariate is redundant in the data-generating mechanism and is adjusted for in some fitted models. The data-generating mechanism of cluster-level covariates remains the same. The general data-generating model is similar:
\begin{align*}
g(\expect[Y_{ij}|\gamma_{i}]) &= \beta^0 + \beta^{\text{person,1}} X^{\text{person,1}}_{ij} + \beta^{\text{person,2}} X^{\text{person,2}}_{ij} + \beta^{\text{person,3}} X^{\text{person,3}}_{ij} \\
&\quad+ \beta^{\text{cluster}} X^{\text{cluster,1}}_i + \gamma_{i}
\end{align*}
with $\beta^0=0$ and the link function $g$ being the logistic link for binary outcomes and the log link for count outcomes. We consider two settings of this general model in terms of which variables are predictive of the outcome. These settings are counterparts of the two settings in the first simulation.
\begin{enumerate}[(A)]
	\addtocounter{enumi}{1} \item The person-level covariates: $\beta^{\text{person,1}}=0.7$, $\beta^{\text{person,2}}=-0.75$, $\beta^{\text{person,3}}=-1$, $\beta^{\text{cluster}}=0$;
	\addtocounter{enumi}{1} \item The person- and cluster-level covariates: $\beta^{\text{person,1}}=0.7$, $\beta^{\text{person,2}}=-0.75$, $\beta^{\text{person,3}}=-1$, $\beta^{\text{cluster}}=0.8$.
\end{enumerate}

The statistical procedures being considered are also similar to the previous simulation with $X^{\text{person,4}}_{ij}$ and $X^{\text{cluster,2}}_{i}$ being the potential redundant covariates. In other words, we consider the models with the following sets of adjusted redundant covariates in addition to the covariates in the data-generating model:
	\begin{enumerate}[(1)]
		\item None;
		\item $X^{\text{person,4}}_{ij}$;
		\item $X^{\text{cluster,2}}_{i}$;
		\item $X^{\text{person,4}}_{ij}$ and $X^{\text{cluster,2}}_{i}$.
	\end{enumerate}
The other two settings of the data-generating model that correspond to A and C in the previous simulation are not of interest, because in the corresponding models, very few person-level covariates are adjusted for, which implies that BW1 and BW2 perform similarly. A particularly interesting combination of the data-generating and fitted models is (D, 4). 
Although primary trial analyses typically adjust for a parsimonious set of covariates known to be associated with the outcome, this scenario may correspond to a secondary analysis that further adjusts for additional variables that may be nonprognostic (e.g., covariates that are not balanced between treatment arms).

{\textbf{\textit{Sensitivity analysis varying simulation parameters}}. To examine the impact of the choice of simulation parameters on the relative performance of the methods, we additionally consider the following alternate settings. First, because many studies may have lower outcome prevalence, we modify the intercept such that the marginal prevalence of the binary outcome is around 10\% {(8.1\% for the scenario we consider)}. Second, we consider smaller covariate effects. The coefficients on the covariates used in our primary simulation reflected values observed in a recently completed parallel group CRT with few clusters \citep{PROUDprotocol,PROUDresult}, but other studies may have smaller coefficients. We therefore set the covariate coefficients in the sensitivity simulation to be half of the values from our second simulation.
For simplicity, these sensitivity simulation scenarios focus on the data-generating mechanism with many covariates (i.e., model~D) and the analytic approach that adjusts for the true set of covariates without redundancy (i.e., model~1).}

\subsection{{PROUD trial}} \label{section: PROUD trial}

The above explorations of covariate-adjusted CRTs in the setting of few clusters were motivated by a parallel-group CRT, the PRimary Care Opioid Use Disorders Treatment (PROUD) Trial \citep{PROUDprotocol,PROUDresult} (ClinicalTrials.gov Identifier: NCT03407638). The PROUD Trial was a pragmatic trial to evaluate a nurse care management intervention for treating opioid use disorder in primary care. Randomization was at the clinic level, stratified by health system: 6 clinics were randomized to the intervention (one per health system) and 6 to continue with usual care (one per health system). The study's powered secondary objective had a patient-level count outcome measure (days of acute care utilization) and a planned GLMM analysis that adjusted for the baseline value of the outcome; pre-planned sensitivity analyses adjusted for a larger number of covariates.

The study sample for patient-level analyses comprised patients with a diagnosis for opioid use disorder (OUD) documented in the EHR during the 3 years pre-randomization. Covariates were assessed during the baseline period  (2 years prior to randomization) and outcomes were assessed over the 2-year follow-up period after randomization. The only exception is that one health care system was randomized 6 months later than the others and its follow-up period is only 18 months after randomization.

We demonstrate the differences in these tests by applying them to patient-level analyses of the PROUD Trial data. For the purpose of this illustration, we consider a binary outcome, the indicator of sustained opioid use disorder (OUD) treatment (having at least 180 days of OUD treatment), and a count outcome, the number of days of acute care utilization. For each outcome, we consider three GLMMs with a clinic-level random intercept that adjust for different sets of covariates: minimally adjusted, adjusted, and fully adjusted. In the minimally adjusted model, following the approach of the main PROUD trial analyses, we adjust for the baseline outcome value; in the adjusted model, we further adjust for a few demographics variables such as gender, age and ethnicity, adding 8 fixed effects at the individual-level; in the fully adjusted model, we additionally adjust for other covariates such as insurance and comorbidity, adding 14 fixed effects at the individual-level. We list details of these covariates in Appendix~\ref{section: PROUD covariate list}. We do not adjust for the stratifying variable, health care system indicator because, otherwise, both BW1 and BW2 DDFs are undefined. In all GLMMs for the count outcome, we set the offset to be the log number of days in the follow-up period.

To compare with the data-generating and fitted models in our simulations, the minimally adjusted model may be similar to (B, 1) in the first simulation; the adjusted model may be similar to (D, 1) in the first simulation, or (D, 4) or (D, 1) in the second simulation and its sensitivity simulation; the fully adjusted model may be similar to but contain more individual-level covariates than (D, 4) in the second simulation or its sensitivity simulation. Notably, there are so many individual-level covariates in the fully adjusted model that BW1 DDF is undefined, while other DDFs are well defined. Even for the adjusted model, BW1 DDF is barely positive and equals one.

\section{Results}

\subsection{Simulation results}

In this section, we present our simulation results, focusing our discussion on a few interesting and representative settings given the large number of scenarios considered (full results are available on GitHub at \url{https://QIU-Hongxiang-David.github.io/small-sample-adjusted-GLMM-CRT/)}. In the three sets of simulations (with 2 included covariates, 4 included covariates, and the sensitivity scenarios), only a small proportion (0.01\%, 0.02\% and 1.71\%, respectively) of model fitting had warnings that indicated slight convergence issues; therefore, these fitted models were included regardless of warnings. Although the primary aim of our simulation studies is to evaluate the methods in terms of type I error rates, we first briefly summarize the bias of the methods, because bias in the estimated treatment effects can contribute to discrepancies between the empirical and nominal type I error rates. We found that for most scenarios, the bias of the treatment effect estimator is small (less than 0.01); the only exception was a setting in the sensitivity simulation where the large bias is caused by an unusually extreme estimate in the simulation, which is likely to be a computation issue or by chance.

\subsubsection{First simulation with at most one person-level covariate} \label{section: results sim1}

\begin{sidewaysfigure}
	\centering
	\includegraphics[scale=.6]{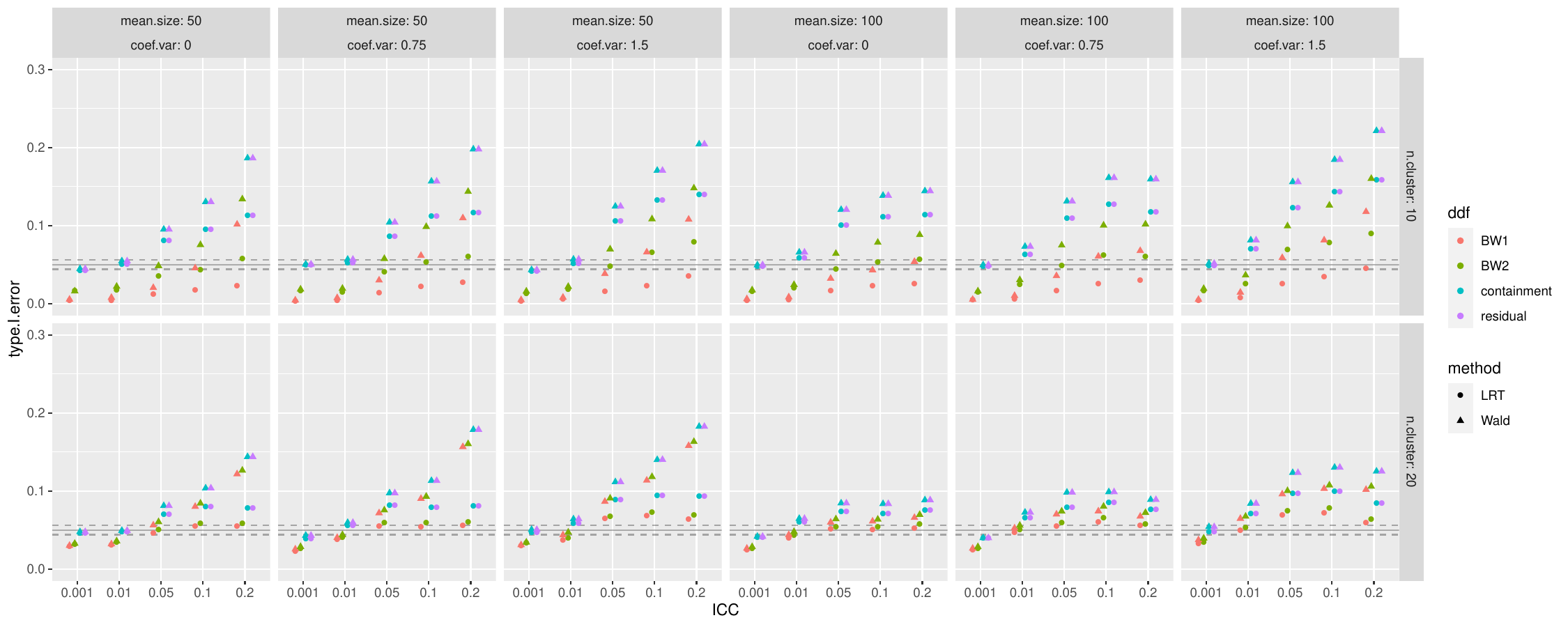}
	\caption{Type I error rates of tests on the treatment effect for data-generating model D and fitted model 1 when the outcome is binary in the first simulation.}
	\label{fig:binary1}
\end{sidewaysfigure}

\begin{sidewaysfigure}
	\centering
	\includegraphics[scale=.6]{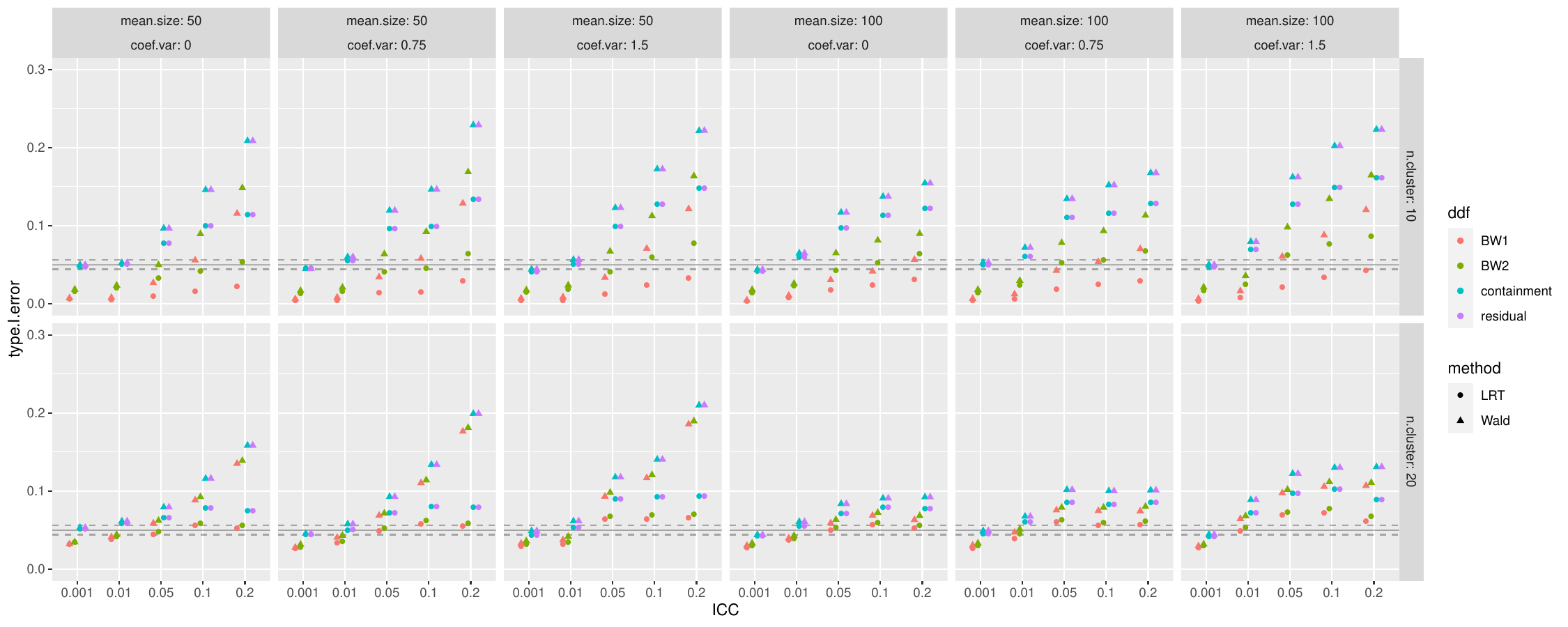}
	\caption{Type I error rates of tests on the treatment effect for data-generating model D and fitted model 1 when the outcome is a count in the first simulation.}
	\label{fig:count1}
\end{sidewaysfigure}

Figures~\ref{fig:binary1} and \ref{fig:count1} present the type I error rates of the tests for binary and count outcomes, under the scenario with data-generating model D and fitted model 1 with a single cluster-level covariate (e..g, stratifying variable) and a single person-level covariate (e.g., the baseline outcome). The relative performance of the methods is similar in the other scenarios considered.

In this scenario, the relative performance of the different methods is similar across the binary and count outcomes. With the same DDF, LRT appears to be generally more conservative than Wald test. All tests appear to be less conservative when the cluster sizes are more variable. 

The containment and residual DDFs appear to perform well only when the ICC is low (0.001, mean LMM ICC 0.0018--0.0123 across both outcome types), in which case the observations are approximately independent. For the other scenarios of ICC ($>0.001$), these two DDF have elevated type I error rates ranging from slightly to highly anticonservative. 

For the other values of ICC of 0.01--0.2 (mean LMM ICC 0.0066--0.1369), when the number of clusters is moderate (20), LRT with BW1 or BW2 DDF appears to have type I error rates either closest or sufficiently close to the nominal level. If the ICC is moderate to high (0.05--0.2, mean LMM ICC 0.0304--0.1358) and the number of clusters is small (10), these two tests also perform the best or reasonably well. BW1 DDF appears to be more appropriate when ICC or CV is relatively high. However, the best-performing tests may still have elevated type I error rates.

The scenario with the most variability in which test has the type I error rate that is closest to the nominal one occurs when ICC is relatively low (0.01, mean LMM ICC 0.0073--0.0183) and the number of clusters is small (10). In this setting, the Wald test with BW1 or BW2 DDF appears to perform well in most scenarios, with some notable exceptions: When the mean cluster size is large (100) and CV is large (1.5), LRT with BW1 or BW2 DDF appears to perform the best, although the type I error rates are still elevated; if the data-generating model is C with fitted model 1, i.e., a single cluster-level covariate is adjusted for (e.g., stratification variable), when the mean cluster size is large (100) and CV is moderate (0.75), LRT with BW1 or BW2 DDF also outperforms Wald test.

\subsubsection{Second simulation with three person-level covariates} \label{section: results sim2}

\begin{sidewaysfigure}
	\centering
	\includegraphics[scale=.6]{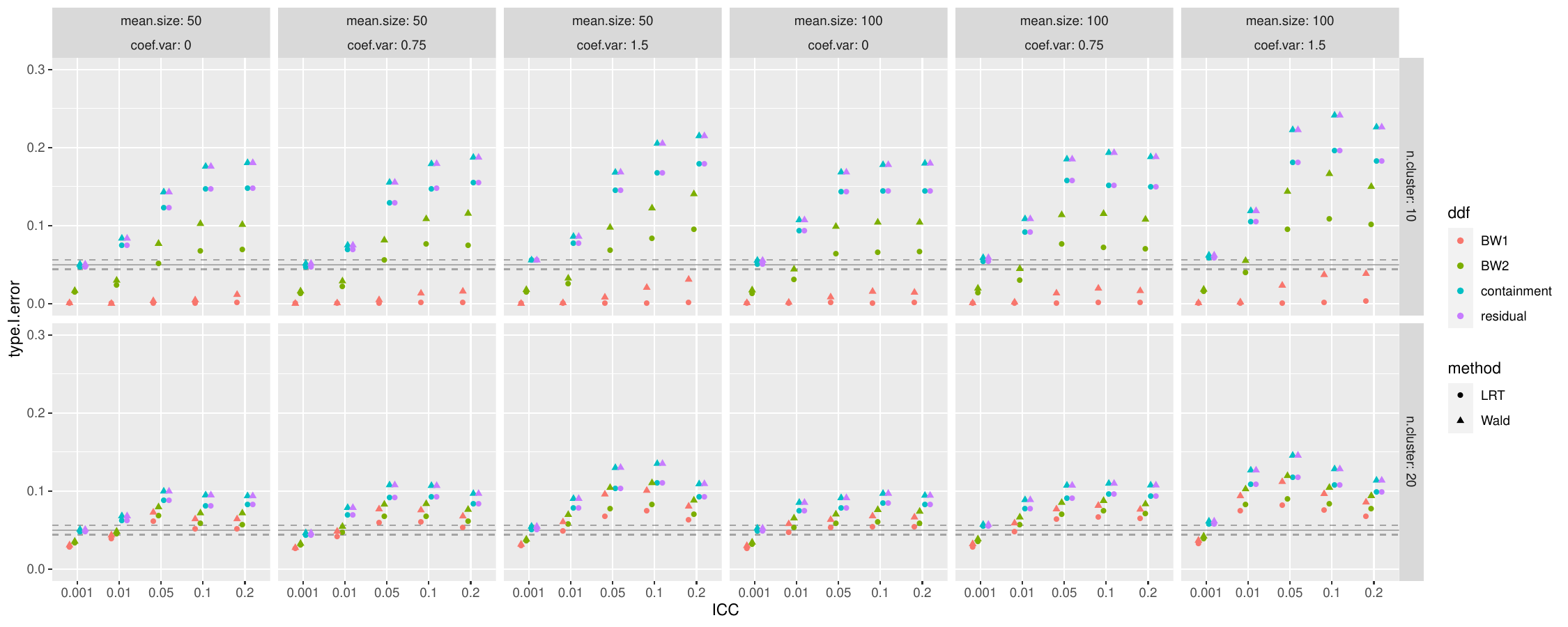}
	\caption{Type I error rates of tests on the treatment effect for data-generating model D and fitted model 4 when the outcome is binary in the second simulation.}
	\label{fig:binary2}
\end{sidewaysfigure}

\begin{sidewaysfigure}
	\centering
	\includegraphics[scale=.6]{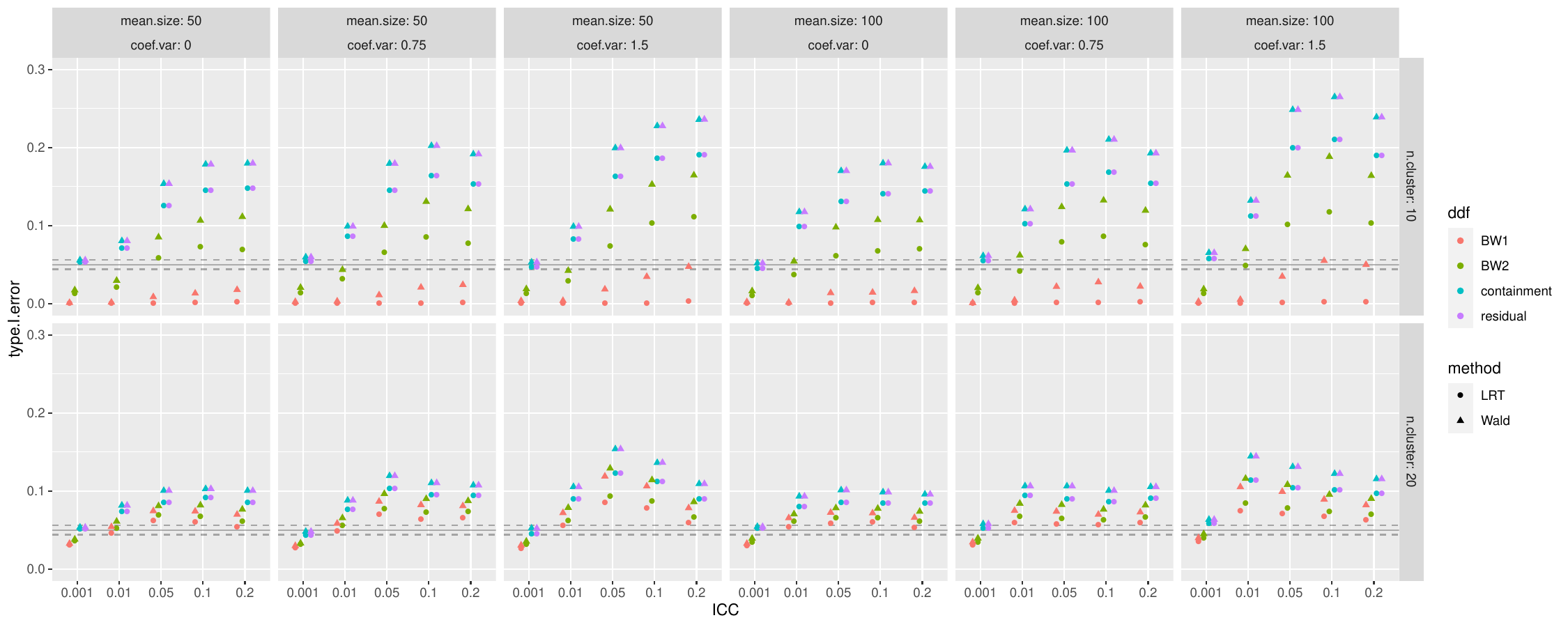}
	\caption{Type I error rates of tests on the treatment effect for data-generating model D and fitted model 4 when the outcome is a count in the second simulation.}
	\label{fig:count2}
\end{sidewaysfigure}

Figures~\ref{fig:binary2} and \ref{fig:count2} present the type I error rates of the tests for binary and count outcomes respectively under data-generating model D and fitted model 4,  which adjusts for 4 person-level covariates (of which 3 are prognostic) and 2 cluster-level covariates (1 prognostic), which corresponds to studies adjusting for several potentially important predictors. As in the first simulation, the relative performance of the different methods is similar across the binary and count outcomes. However, the results are more mixed and inconclusive across scenarios because BW1 DDF may lead to substantially conservative tests, when there are a small number of clusters (10), even in settings of moderate to high ICCs where it achieved good performance in the first simulation setting with at most one person-level covariate.

When ICC is low (0.001, mean LMM ICC 0.0017--0.0189), similarly to the first simulation, residual and containment DDFs still perform well. When the number of clusters is moderate (20), both LRTs with BW1 and BW2 DDF appear to have size closer to the nominal level than other tests for ICC being 0.01, 0.05, 0.1 and 0,2 (mean LMM ICC 0.0032--0.1253). BW1 appears to perform slightly better, but their difference is small.

For the rest of the section, we will focus on the setting with 10 clusters and ICC higher than 0.001, in which case the best-performing method was highly dependent on the scenario.

When ICC is relatively low (0.01, mean LMM ICC 0.0040--0.0223), Wald test with BW2 DDF is usually among the best-performing tests.
This is similar to our finding in the first simulation. When ICC is moderate to high (0.05, 0.1, 0.2, mean LMM ICC 0.0130--0.1240), usually either Wald test with BW1 DDF or LRT with BW2 DDF appears to have size closer to the nominal level than other tests. When the CV of cluster sizes is large (1.5), Wald test with BW1 DDF appears to have type I error closest to the nominal level; when CV is 0 to moderate (0.75), LRT with BW2 DDF appears to perform best. We note that Wald test with BW1 DDF appears to be more conservative than LRT with BW2 DDF when the number of clusters is small (10) and therefore may be preferable if no information is available to determine which test has size closer to the nominal level.

An interesting finding is that when the CV is high (1.5), most tests have inflated type I error rates. This suggests that, with highly variable cluster sizes and more than a few covariates to be adjusted for, it is often difficult or even impossible to identify a test with acceptable type I error rates.

{\textbf{\textit{Sensitivity analysis varying the simulation parameters}}. Figures~\ref{fig:binary3} and \ref{fig:count3} present the type I error rates of the tests for binary and count outcomes respectively under data-generating model~D and fitted model~1. The results appear similar to the second simulation in Section~\ref{section: results sim2} with the following differences when the number of clusters is small (10). When the ICC is relatively low (0.01, mean LMM ICC 0.0056--0.0131) and the mean cluster size is small (50), both between-within DDFs appear too conservative while containment and residual DDFs appear to have size closest to the nominal level. With a high CV of cluster sizes (1.5) and a relatively high ICC (0.1 and 0.2, mean LMM ICC 0.0287--0.0797), LRT with BW1 DDF appears to have size closest to the nominal level. These differences are likely to be caused by the much lower prevalence in the third simulation. Therefore, we conclude that it is generally difficult to find a test \emph{a priori} with approximately valid type I error rates if the number of clusters is small, the cluster sizes vary substantially, the ICC is unknown, and more than a few covariates are adjusted for.}

\subsection{PROUD trial} \label{section: PROUD trial results}

The PROUD trial data set consists of 1,988 patients. The mean cluster (clinic) size is 166 and the CV of cluster sizes is 0.7. Of these patients, 5.0\% had sustained OUD treatment; the average number of days of acute care utilization is 5.2. We next present our analysis results for the two outcomes separately.

\subsubsection{Sustained OUD treatment (binary outcome)}

The proportions of patients having sustained OUD treatment in the intervention and control arms are 3.3\% and 6.6\%, respectively. The corresponding na\"ive estimate of the log odds ratio is 0.74, without accounting for the clustered structure and baseline covariates. These proportions in the baseline period in the intervention and control arms are 3.2\% and 6.9\%, respectively. Thus, we expect that adjusting for baseline covariates would significantly improve estimation of the intervention effect. According to the fitted GLMMs, the minimally adjusted, adjusted, and fully adjusted log odds ratios are -0.52 (SE 0.27), -0.56 (SE 0.27), and -0.55 (SE 0.28), respectively. As we expected, adjusting for the baseline outcome leads to a somewhat different estimate of log odds ratio from the na\"ive estimate, while adjusting for other covarites does not considerably change the estimate. The estimated ICCs from the corresponding models are all almost zero. According to our simulation results, such almost negligible ICC suggests that residual and containment DDFs may be appropriate.

Table~\ref{table: PROUD binary p-value} presents p-values of the testing methods we consider. Only four p-values are barely below 0.05: the Wald and LRT tests with residual and containment DDFs in the adjusted model. Tests with residual and containment DDFs all have p-values close to 0.05, while tests with BW1 and BW2 DDFs have p-values greater than 0.05. BW1 DDF appears to be too conservative for the adjusted model, while BW2 DDF appears reasonably more conservative than residual and containment DDFs. These results show that if a conventional level of significance of 0.05 is used, different choices for the analytic method would lead to different conclusions on whether to reject the null hypothesis.
If the p-values are only interpreted as a measure of strength of evidence, these two approaches lead to somewhat similar conclusions.

\subsubsection{Days of acute care utilization (count outcome)}

The mean number of days of acute care utilization per year in the intervention and control arms are 3.0 and 2.2, respectively. The corresponding na\"ive estimate of log rate ratio is -0.30, without accounting for the clustered structure and baseline covariates. These mean numbers in the intervention and control arms are 9.2 and 9.7, respectively. According to the fitted GLMMs, the minimally adjusted, adjusted and fully adjusted log rate ratios are 0.15 (SE 0.37), 0.15 (SE 0.38) and 0.20 (SE 0.40) respectively. Adjusting for the baseline and accounting for the length of follow-up lead to estimates with a different sign, while adjusting for other covarites does not considerably change the estimate. Since the standard errors are large compared to the point estimates, the change in the estimates might not be meaningful. The estimated ICCs from the corresponding models are 0.07, 0.07 and 0.08 respectively. To estimate the LMM ICC while accounting for the length of follow-up, we fit LMMs with the same covariates and the outcome scaled by the length of follow-up. The estimated LMM ICCs are all about 0.012. Thus, the ICC is moderate.

Table~\ref{table: PROUD count p-value} presents p-values of the testing methods we consider under the minimally adjusted, adjusted and fully adjusted models. Although none of them is significant at significance level 0.05, their strengths of evidence differ with a similar pattern to that observed in the binary outcome analysis.

\section{Discussion}

In this study, we examined the performance of routinely available small-sample corrected statistical tests for GLMMs of binary and count outcomes with covariate adjustment under a CRT setting. Through simulation studies, we found that when the number of clusters is moderate (20), with a non-negligible ICC ($\geq 0.01$), LRT with BW1 or BW2 DDF have type I error rates close to the nominal level across settings of covariate adjustment. These tests also perform well when the number of clusters is small (10) and the ICC is at least moderate, provided only a few covariates are adjusted for (though an exception occurs when the CV of cluster sizes is high -- e.g., 1.5 -- in which case all of the methods considered had inflated type I error rates).
In contrast, with few clusters, if even a moderate number of covariates are adjusted for (e.g., 4--6), the performance of the tests varies dramatically across scenarios.

Given this high degree of variability in the optimal choice of the procedure with few clusters, we recommend limiting the number of covariates being adjusted for in the primary analysis (e.g., to at most two degrees of freedom), regardless of whether the covariates are at the cluster level or the person level. Our results further suggest that LRT with BW1 or BW2 DDF performs reasonably well across a broad range of settings. With additional covariate adjustment, knowledge of the number of clusters and the CV of the cluster sizes can be used to inform the selection of the best-performing test in that scenario. However, the true value of the ICC is often not known, and estimates from prior studies or baseline data can be highly variable. Thus, it may be impractical to pick an appropriate test based on which scenario the trial is in. In this case, we recommend that analyses adjusting for more covariates be considered as secondary or sensitivity analyses.

There are several limitations of the present study and remaining gaps in knowledge. First, we restricted to binary and count outcomes and did not consider continuous outcomes and LMM. Despite rich literature on tests for LMM with few clusters in CRTs, the optimal choice of method in covariate-adjusted analyses remains an open question. Second, although we considered a wide range of realistic scenarios, our results might not generalize to scenarios beyond those considered, such as when the person-level covariates are clustered or the randomization is stratified. Third, as noted by the literature, there may be multiple sources of error in inference based on GLMM, including computation algorithms \citep{Elff2020}, which could impact the relative performance of methods. Finally, we only considered scenarios where person-level covariates are independent of the cluster-level treatment assignment. Such scenarios would occur if analyses are restricted to a cohort prior to randomization. However, some CRTs may be at risk of systematic bias, which can be caused, for example, by recruiting people after randomization \citep{bobb2020,li2021}. Our findings may not translate to this setting.
	
In this study, we did not consider Kenward-Roger's or Satterthwaite's methods to compute the DDF. A prior study found that these methods were outperformed by between-within methods for logistic GLMMs \citep{li2015}. In addition, although these DDFs are commonly used in LMMs, to our best knowledge, the corresponding methods for GLMMs have not been routinely implemented (in common statistical packages, they are available only in SAS). In contrast, the other DDFs can even be calculated by hand. We therefore focused on available methods that can be computed easily across a range of software programs.

Another interesting finding is that, in the cases where LRT with BW1 or BW2 DDF perform well, LRTs appear to perform better than Wald tests with the same model and same DDF. This finding differs from prior studies that suggested that LRTs may be unreliable with a small number of clusters \citep{li2015,bolker2009,vonesh2014}. However, as noted in the literature, LRT may be adequate when the total number of individuals is large compared with the number of fixed effect levels and the number of clusters \citep{bolker2009}, which in fact often holds for CRTs. Our simulation results suggest that LRT may outperform Wald tests in such settings.

In summary, this work examined the performance of small sample-corrected tests for GLMM of the treatment effect in a cluster-randomized trial setting, after adjusting for covariates. Our simulations considered a breadth of realistic scenarios that varied the sample size (including the number of clusters, mean cluster size, and variability of cluster sizes), the degree of within-cluster correlation, and the number of prognostic covariates. We also considered different analytic strategies for covariate adjustment, including adjusting for just a few versus a larger number of covariates. We found that the relative performance of the testing methods was highly variable across different scenarios. Adjusting for only a few covariates and using likelihood ratio tests with a between-within degree of freedom appears to perform well across many of the settings we considered. Nonetheless, because no method performs uniformly well and the choice of optimal method could vary quite substantially in some scenarios, statisticians are encouraged to conduct a simulation study tailored to their particular trial setting before specifying the choice of testing procedure in the statistical analysis plan.

\section*{Data availability statement}

Codes to generate the simulated datasets used in the simulation study are openly available at \url{https://github.com/QIU-Hongxiang-David/small-sample-adjusted-GLMM-CRT}. Electronic health record data used in the application are not publicly available due to privacy or ethical restrictions.

\bibliographystyle{abbrvnat}
\bibliography{references}

\newcommand{\noop}[1]{}
\begin{thebibliography}{24}
\providecommand{\natexlab}[1]{#1}
\providecommand{\url}[1]{\texttt{#1}}
\expandafter\ifx\csname urlstyle\endcsname\relax
  \providecommand{\doi}[1]{doi: #1}\else
  \providecommand{\doi}{doi: \begingroup \urlstyle{rm}\Url}\fi

\bibitem[Bates et~al.(2015)Bates, M{\"a}chler, Bolker, and Walker]{lme4package}
D.~Bates, M.~M{\"a}chler, B.~Bolker, and S.~Walker.
\newblock Fitting linear mixed-effects models using {lme4}.
\newblock \emph{Journal of Statistical Software}, 67\penalty0 (1):\penalty0
  1--48, 2015.
\newblock \doi{10.18637/jss.v067.i01}.

\bibitem[Bobb et~al.(2020)Bobb, Qiu, Matthews, McCormack, and
  Bradley]{bobb2020}
J.~F. Bobb, H.~Qiu, A.~G. Matthews, J.~McCormack, and K.~A. Bradley.
\newblock Addressing identification bias in the design and analysis of
  cluster-randomized pragmatic trials: a case study.
\newblock \emph{Trials}, 21\penalty0 (1):\penalty0 1--12, 2020.

\bibitem[Bolker et~al.(2009)Bolker, Brooks, Clark, Geange, Poulsen, Stevens,
  and White]{bolker2009}
B.~M. Bolker, M.~E. Brooks, C.~J. Clark, S.~W. Geange, J.~R. Poulsen, M.~H.~H.
  Stevens, and J.-S.~S. White.
\newblock Generalized linear mixed models: a practical guide for ecology and
  evolution.
\newblock \emph{Trends in ecology \& evolution}, 24\penalty0 (3):\penalty0
  127--135, 2009.

\bibitem[Campbell et~al.(2021)Campbell, Saxon, Boudreau, Wartko, Bobb, Lee,
  Matthews, McCormack, Liu, Addis, et~al.]{PROUDprotocol}
C.~I. Campbell, A.~J. Saxon, D.~M. Boudreau, P.~D. Wartko, J.~F. Bobb, A.~K.
  Lee, A.~G. Matthews, J.~McCormack, D.~S. Liu, M.~Addis, et~al.
\newblock Primary care opioid use disorders treatment (proud) trial protocol: a
  pragmatic, cluster-randomized implementation trial in primary care for opioid
  use disorder treatment.
\newblock \emph{Addiction science \& clinical practice}, 16\penalty0
  (1):\penalty0 1--15, 2021.

\bibitem[Donner and Koval(1980)]{donner1980}
A.~Donner and J.~J. Koval.
\newblock The estimation of intraclass correlation in the analysis of family
  data.
\newblock \emph{Biometrics}, pages 19--25, 1980.

\bibitem[Eldridge et~al.(2004)Eldridge, Ashby, Feder, Rudnicka, and
  Ukoumunne]{eldridge2004}
S.~M. Eldridge, D.~Ashby, G.~S. Feder, A.~R. Rudnicka, and O.~C. Ukoumunne.
\newblock Lessons for cluster randomized trials in the twenty-first century: a
  systematic review of trials in primary care.
\newblock \emph{Clinical trials}, 1\penalty0 (1):\penalty0 80--90, 2004.

\bibitem[Elff et~al.(2021)Elff, Heisig, Schaeffer, and Shikano]{Elff2020}
M.~Elff, J.~P. Heisig, M.~Schaeffer, and S.~Shikano.
\newblock Multilevel analysis with few clusters: improving likelihood-based
  methods to provide unbiased estimates and accurate inference.
\newblock \emph{British Journal of Political Science}, 51\penalty0
  (1):\penalty0 412--426, 2021.

\bibitem[Elixhauser et~al.(1998)Elixhauser, Steiner, Harris, and
  Coffey]{elixhauser1998}
A.~Elixhauser, C.~Steiner, D.~R. Harris, and R.~M. Coffey.
\newblock Comorbidity measures for use with administrative data.
\newblock \emph{Medical care}, pages 8--27, 1998.

\bibitem[Kahan et~al.(2014)Kahan, Jairath, Dor{\'e}, and Morris]{kahan2014}
B.~C. Kahan, V.~Jairath, C.~J. Dor{\'e}, and T.~P. Morris.
\newblock The risks and rewards of covariate adjustment in randomized trials:
  an assessment of 12 outcomes from 8 studies.
\newblock \emph{Trials}, 15\penalty0 (1):\penalty0 139, 2014.

\bibitem[Kahan et~al.(2016)Kahan, Forbes, Ali, Jairath, Bremner, Harhay,
  Hooper, Wright, Eldridge, and Leyrat]{kahan2016}
B.~C. Kahan, G.~Forbes, Y.~Ali, V.~Jairath, S.~Bremner, M.~O. Harhay,
  R.~Hooper, N.~Wright, S.~M. Eldridge, and C.~Leyrat.
\newblock Increased risk of type i errors in cluster randomised trials with
  small or medium numbers of clusters: a review, reanalysis, and simulation
  study.
\newblock \emph{Trials}, 17\penalty0 (1):\penalty0 438, 2016.

\bibitem[Leyrat et~al.(2018)Leyrat, Morgan, Leurent, and Kahan]{leyrat2018}
C.~Leyrat, K.~E. Morgan, B.~Leurent, and B.~C. Kahan.
\newblock Cluster randomized trials with a small number of clusters: which
  analyses should be used?
\newblock \emph{International journal of epidemiology}, 47\penalty0
  (1):\penalty0 321--331, 2018.

\bibitem[Li et~al.(2022)Li, Tian, Bobb, Papadogeorgou, and Li]{li2021}
F.~Li, Z.~Tian, J.~Bobb, G.~Papadogeorgou, and F.~Li.
\newblock Clarifying selection bias in cluster randomized trials.
\newblock \emph{Clinical Trials}, 19\penalty0 (1):\penalty0 33--41, 2022.

\bibitem[Li and Redden(2015)]{li2015}
P.~Li and D.~T. Redden.
\newblock Comparing denominator degrees of freedom approximations for the
  generalized linear mixed model in analyzing binary outcome in small sample
  cluster-randomized trials.
\newblock \emph{BMC medical research methodology}, 15\penalty0 (1):\penalty0
  38, 2015.

\bibitem[McNeish and Stapleton(2016)]{mcneish2016}
D.~McNeish and L.~M. Stapleton.
\newblock Modeling clustered data with very few clusters.
\newblock \emph{Multivariate behavioral research}, 51\penalty0 (4):\penalty0
  495--518, 2016.

\bibitem[Nakagawa et~al.(2017)Nakagawa, Johnson, and
  Schielzeth]{nakagawa2017ICC}
S.~Nakagawa, P.~C. Johnson, and H.~Schielzeth.
\newblock The coefficient of determination $r^2$ and intra-class correlation
  coefficient from generalized linear mixed-effects models revisited and
  expanded.
\newblock \emph{Journal of the Royal Society Interface}, 14\penalty0
  (134):\penalty0 20170213, 2017.

\bibitem[Pinheiro and Bates(2006)]{pinheiro2006}
J.~Pinheiro and D.~Bates.
\newblock \emph{Mixed-effects models in S and S-PLUS}.
\newblock Springer Science \& Business Media, 2006.

\bibitem[Roberts and Torgerson(1999)]{roberts1999}
C.~Roberts and D.~J. Torgerson.
\newblock Baseline imbalance in randomised controlled trials.
\newblock \emph{Bmj}, 319\penalty0 (7203):\penalty0 185, 1999.

\bibitem[Schaalje et~al.(2002)Schaalje, McBride, and Fellingham]{schaalje2002}
G.~B. Schaalje, J.~B. McBride, and G.~W. Fellingham.
\newblock Adequacy of approximations to distributions of test statistics in
  complex mixed linear models.
\newblock \emph{Journal of Agricultural, Biological, and Environmental
  Statistics}, 7\penalty0 (4):\penalty0 512, 2002.

\bibitem[Schluchter and Elashoff(1990)]{schluchter1990}
M.~D. Schluchter and J.~T. Elashoff.
\newblock Small-sample adjustments to tests with unbalanced repeated measures
  assuming several covariance structures.
\newblock \emph{Journal of Statistical Computation and Simulation}, 37\penalty0
  (1-2):\penalty0 69--87, 1990.

\bibitem[Stanish and Taylor(1983)]{stanish1983}
W.~M. Stanish and N.~Taylor.
\newblock Estimation of the intraclass correlation coefficient for the analysis
  of covariance model.
\newblock \emph{The American Statistician}, 37\penalty0 (3):\penalty0 221--224,
  1983.

\bibitem[Thompson et~al.(2022)Thompson, Leyrat, Fielding, and
  Hayes]{thompson2022}
J.~A. Thompson, C.~Leyrat, K.~L. Fielding, and R.~J. Hayes.
\newblock Cluster randomised trials with a binary outcome and a small number of
  clusters: comparison of individual and cluster level analysis method.
\newblock \emph{BMC Medical Research Methodology}, 22\penalty0 (1):\penalty0
  1--15, 2022.

\bibitem[Vonesh(2014)]{vonesh2014}
E.~F. Vonesh.
\newblock \emph{Generalized linear and nonlinear models for correlated data:
  theory and applications using SAS}.
\newblock SAS Institute, 2014.

\bibitem[Wartko et~al.(2023)Wartko, Bobb, Boudreau, Matthews, McCormack, Lee,
  Qiu, Yu, Hyun, Idu, Campbell, Saxon, Liu, Altschuler, Samet, Labelle,
  Zare-Mehrjerdi, Stotts, Braciszewski, Murphy, Dryden, Arnsten, Cunningham,
  Horigian, Szapocznik, Glass, Caldeiro, Phillips, Shea, Bart, Schwartz,
  McNeely, Liebschutz, Tsui, Merrill, Lapham, Addis, and Bradley]{PROUDresult}
P.~D. Wartko, J.~F. Bobb, D.~M. Boudreau, A.~G. Matthews, J.~McCormack, A.~K.
  Lee, H.~Qiu, O.~Yu, N.~Hyun, A.~E. Idu, C.~I. Campbell, A.~J. Saxon, D.~S.
  Liu, A.~Altschuler, J.~H. Samet, C.~T. Labelle, M.~Zare-Mehrjerdi, A.~L.
  Stotts, J.~M. Braciszewski, M.~T. Murphy, D.~Dryden, J.~H. Arnsten, C.~O.
  Cunningham, V.~E. Horigian, J.~Szapocznik, J.~E. Glass, R.~M. Caldeiro, R.~C.
  Phillips, M.~Shea, G.~Bart, R.~P. Schwartz, J.~McNeely, J.~M. Liebschutz,
  J.~I. Tsui, J.~O. Merrill, G.~T. Lapham, M.~Addis, and K.~A. Bradley.
\newblock The primary care opioid use disorders treatment (proud) trial: a
  cluster-randomized implementation trial of nurse care management for opioid
  use disorder treatment.
\newblock \emph{JAMA Internal Medicine}, 2023.
\newblock ISSN 2168-6106.
\newblock \doi{10.1001/jamainternmed.2023.5701}.

\bibitem[Wright et~al.(2015)Wright, Ivers, Eldridge, Taljaard, and
  Bremner]{wright2015}
N.~Wright, N.~Ivers, S.~Eldridge, M.~Taljaard, and S.~Bremner.
\newblock A review of the use of covariates in cluster randomized trials
  uncovers marked discrepancies between guidance and practice.
\newblock \emph{Journal of clinical epidemiology}, 68\penalty0 (6):\penalty0
  603--609, 2015.

\end{thebibliography}

\appendix

\section{Type I error rates in the sensitivity simulation}

\begin{sidewaysfigure}[h!]
	\centering
	\includegraphics[scale=.6]{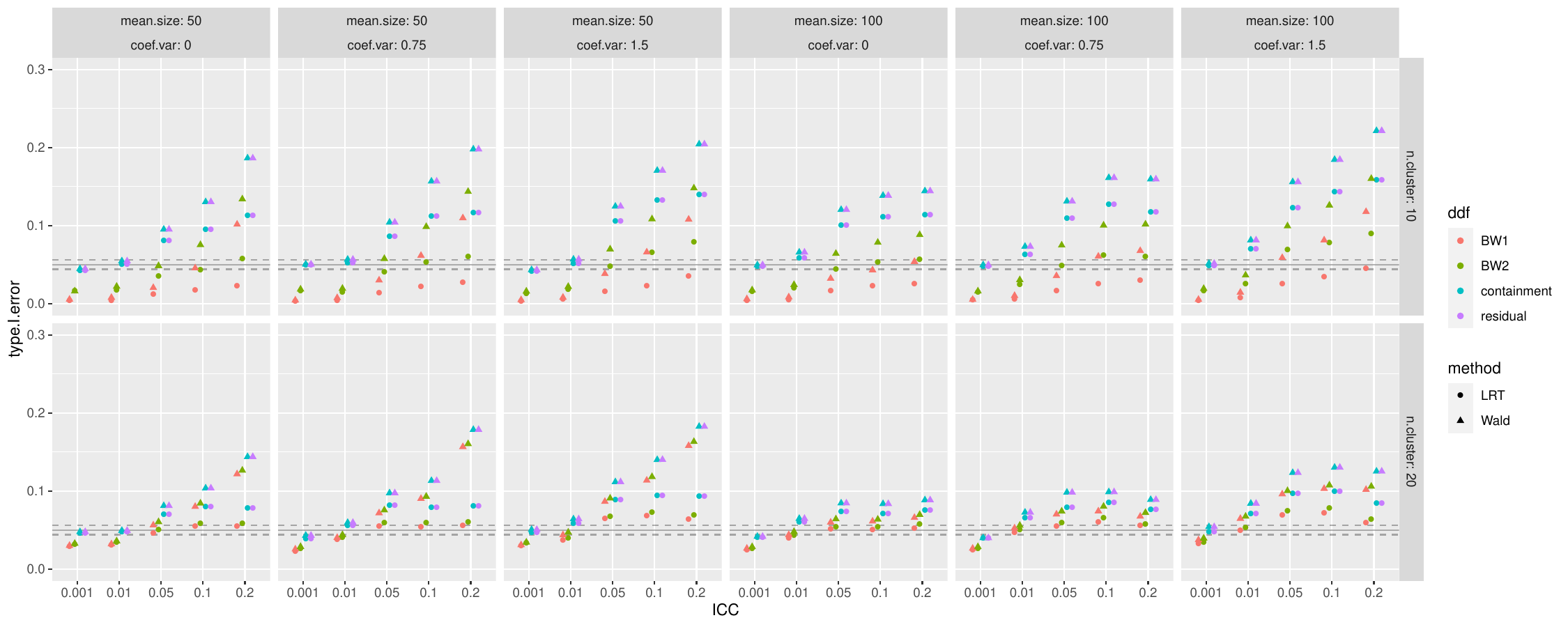}
	\caption{Type I error rates of tests on the treatment effect for data-generating model D and fitted model 1 when the outcome is binary in the third simulation.}
	\label{fig:binary3}
\end{sidewaysfigure}

\begin{sidewaysfigure}[h!]
	\centering
	\includegraphics[scale=.6]{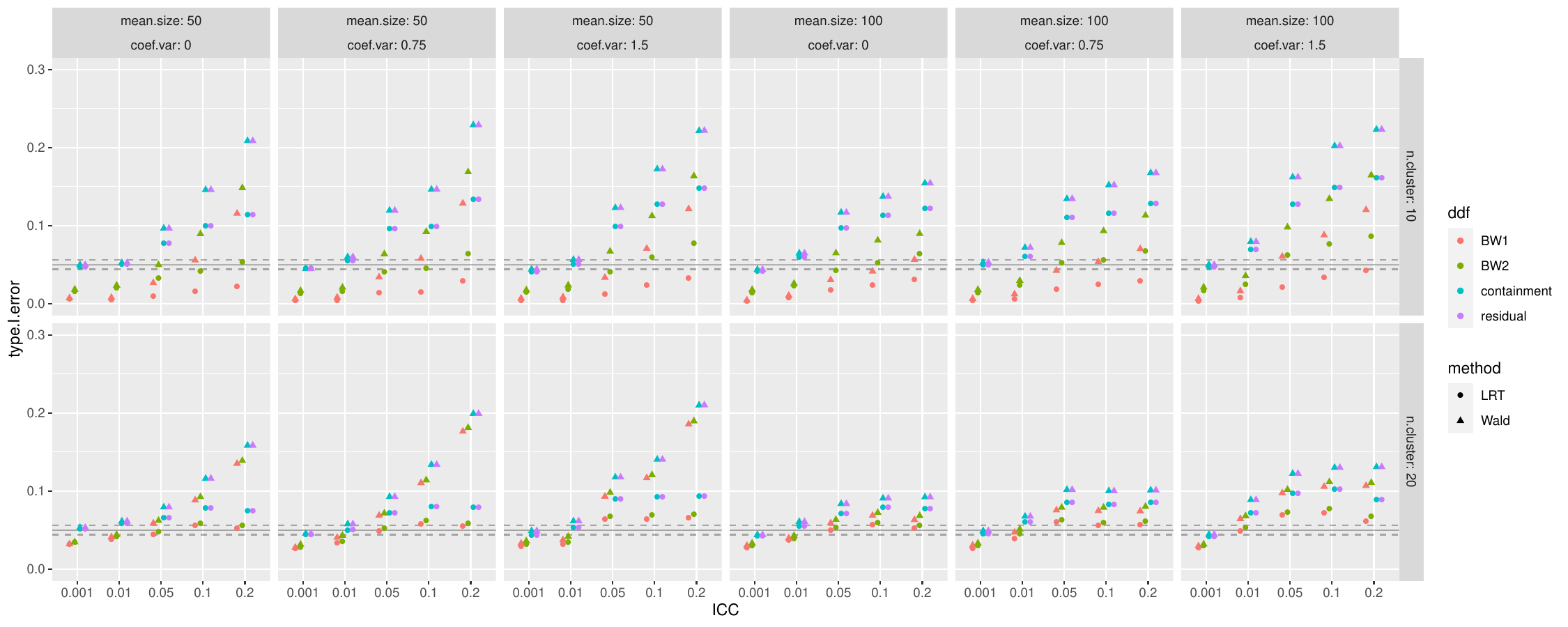}
	\caption{Type I error rates of tests on the treatment effect for data-generating model D and fitted model 1 when the outcome is a count in the third simulation.}
	\label{fig:count3}
\end{sidewaysfigure}

\newpage

\begin{table}[h!]
	\centering
	\caption{P-values of various two-sided tests of the treatment effect on the binary outcome under the adjusted and fully adjusted models. DDF: denominator degree of freedom. BW: between-within. BW1 DDF is undefined for the fully adjusted model.}
	\label{table: PROUD binary p-value}
	\begin{tabular}{l|l||r|r|r}
		\hline
		DDF & Testing method & Minimally adjusted & Adjusted & Fully adjusted \\
		\hline \hline
		residual \rule{0pt}{3ex} & Wald & $0.054$ & $0.042$ & $0.051$ \\
		& LRT & $0.051$ & $0.039$ & $0.048$ \\
		containment \rule{0pt}{4ex} & Wald & $0.054$ & $0.042$ & $0.051$ \\
		& LRT & $0.051$ & $0.039$ & $0.048$ \\
		BW1 \rule{0pt}{4ex} & Wald & $0.086$ & $0.290$ & --- \\
		& LRT & $0.082$ & $0.286$ & --- \\
		BW2 \rule{0pt}{4ex} & Wald & $0.083$ & $0.069$ & $0.080$ \\
		& LRT & $0.079$ & $0.065$ & $0.076$ \\
		\hline
	\end{tabular}
\end{table}

\begin{table}[h!]
	\centering
	\caption{P-values of various two-sided tests of the treatment effect on the count outcome under the adjusted and fully adjusted models. DDF: denominator degree of freedom. BW: between-within. BW1 DDF is undefined for the fully adjusted model.}
	\label{table: PROUD count p-value}
	\begin{tabular}{l|l||r|r|r}
		\hline
		DDF & Testing method & Minimally adjusted & Adjusted & Fully adjusted \\
		\hline \hline
		residual \rule{0pt}{3ex} & Wald & $0.691$ & $0.693$ & $0.627$ \\
		& LRT & $0.692$ & $0.693$ & $0.627$ \\
		containment \rule{0pt}{4ex} & Wald & $0.691$ & $0.693$ & $0.627$ \\
		& LRT & $0.692$ & $0.693$ & $0.627$ \\
		BW1 \rule{0pt}{4ex} & Wald & $0.700$ & $0.760$ & --- \\
		& LRT & $0.701$ & $0.760$ & --- \\
		BW2 \rule{0pt}{4ex} & Wald & $0.700$ & $0.701$ & $0.637$ \\
		& LRT & $0.700$ & $0.701$ & $0.637$ \\
		\hline
	\end{tabular}
\end{table}

\section{Covariates being adjusted for in PROUD data analysis} \label{section: PROUD covariate list}

In the minimally adjusted model, we adjust for the baseline outcome. For the binary outcome (indicator of sustained OUD treatment), we use the continuous variable, number of days of OUD medication treatment in the baseline period, as the baseline outcome to be adjusted for. We do not use a dichotomized version similar to the binary outcome to preserve more information in covariates. In the minimally adjusted model, there are three fixed effects: intercept, treatment arm (cluster-level), and baseline outcome (individual-level).

In the adjusted model, we adjust for the following covariates in addition to the baseline outcome:
\begin{itemize}
	\item age at randomization, including both linear and quadratic terms;
	\item gender;
	\item race/ethnicity with six categories: White, Asian, Black, Hispanic, Other, and Unknown.
\end{itemize}
Thus, there are eight more fixed effects associated with individual-level covariates compared to the minimally adjusted model.

In the fully adjusted model, we further adjust for the following covariates:
\begin{itemize}
	\item three neighborhood-level measures capturing socioeconomic status: median household income, percent below federal poverty line, and percent unemployed;
	\item four insurance status indicators: Medicaid,
Medicare, others, and uninsured;
	\item number of days of OUD medication treatment in the baseline period\footnote{For the binary outcome, this covariate is the baseline outcome and thus is excluded from this list.};
	\item number of days with documented OUD diagnosis;
	\item four baseline comorbidity variables: indicator of alcohol use disorder diagnosis; indicator of other substance use disorder diagnosis; a weighted summary score of other comorbidities (Elixhauser index \citep{elixhauser1998})
	\item indicator of housing instability in the baseline period.
\end{itemize}
Thus, there are 13 (for binary outcome) or 14 (for count outcome) fixed effects associated with individual-level covariates compared to the adjusted model.

\end{document}